\documentclass[runningheads]{llncs}

\usepackage[utf8]{inputenc}
\usepackage{xspace}

\usepackage{booktabs} 
\usepackage{microtype}
\usepackage{graphicx}
\usepackage{booktabs} 
\usepackage{verbatim}
\usepackage{esint}
\usepackage{varwidth}
\usepackage{url}
\usepackage{nicefrac}
\usepackage{graphicx}
\usepackage{float}
\usepackage[]{hyperref}
\usepackage{bm}
\usepackage{tikz}
\usepackage{setspace}
\usepackage{epstopdf}
\usepackage{etoolbox}
\usepackage{subcaption}

\newcommand{\one}{({\em i}\/)\xspace}
\newcommand{\two}{({\em ii}\/)\xspace}

\def\eg{\emph{e.g., }\xspace}

\def\ie{\emph{i.e.,}\xspace}

\def\vs{\emph{vs.}\xspace}

\newcommand{\pb}[1]{\vspace{0.75ex}\noindent{\bf \em #1}\hspace*{.3em}}

\newcommand\gareth[1]{\textbf{\textcolor{blue}{GT: #1}}}
\newcommand\ag[1]{\textbf{\textcolor{orange}{AG: #1}}}
\newcommand\zg[1]{\textbf{\textcolor{red}{ZG: #1}}}
\newcommand\ar[1]{\textbf{\textcolor{purple}{AR: #1}}}

\newcommand\oliver[1]{\textcolor{blue}{#1}}

\renewcommand\gareth[1]{} \renewcommand\ag[1]{} \renewcommand\zg[1]{} \renewcommand\ar[1]{}

\renewcommand\oliver[1]{#1}

\title{Dissecting the Workload of a Major Adult Video Portal\thanks{Preprint - To be published at PAM 2020}}

\author{
Andreas Grammenos\inst{1,3} \and
Aravindh Raman\inst{2,3} \and
Timm B{\"o}ttger\inst{3} \and 
Zafar Gilani\inst{3} \and
Gareth Tyson\inst{3}
}

\authorrunning{Grammenos et al. (preprint)}

\institute{
University of Cambridge \and
King's College London \and
Queen Mary University of London
}

\begin{document}

\maketitle

\begin{abstract}
Adult content constitutes a major source of Internet traffic. 
As with many other platforms, these sites are incentivized to engage users and maintain them on the site. 
This engagement (\eg through recommendations) shapes the journeys taken through such sites. 
Using data from a large content delivery network, we explore session journeys within an adult website. 
We take two perspectives. 
We first inspect the corpus available on these platforms. 
Following this, we investigate the session access patterns. 
We make a number of observations that could be exploited for optimizing delivery, \eg that users often skip within video streams. 

\end{abstract}

\section{Introduction}

The Internet has evolved from a largely web-oriented infrastructure to a massively distributed content delivery system \cite{arbor}. 
Video content has become particularly popular, and we have therefore seen a range of studies investigating the usage and access patterns of major portals, \eg user-generated content (UGC) \cite{raman2018facebook,YouTube}, video on demand (VoD) \cite{yu2006understanding}, Internet TV (IPTV) \cite{cha2008watching} and catch-up TV \cite{Abrahamsson2012popularity,NRS13www}. 
A particularly prevalent form of online video is that of \emph{adult content}, \ie pornographic material~
\cite{ahmed2016internet}.
In the last five years there has been a surge of research activity in this space, attempting to characterize the content corpus of sites~\cite{tyson2016measurements,tyson2013demystifying}, the workload of sites~\cite{yu2019comparative,ahmed2016internet} and the use of adult social networks~\cite{tyson2015people}. 
Despite this, we still lack the breadth and depth of understanding common to many other aspects of online video delivery. 
Due to the paucity of data, there is a particular lack of understanding related to the unique workload that such websites place on the infrastructure of a Content Delivery Network (CDN). 
Particularly, there has been limited work exploring the per-session content request patterns on these portals. 
Thus, in this paper, we present a large-scale analysis of access patterns for a major adult website, with a focus on understanding how individual viewer decisions (or ``journeys'') impact the workload observed. 
To achieve this, we bring together two key datasets. 
We have gathered data from a large CDN, covering 1 hour of access logs for resources hosted served by the site.
This covers 20.08M access records, 62K users and 3.28TB of exchanged data.
Although this offers fine-grained insight into content request patterns, alone is it insufficient. 
This is because modern adult websites also consist of a large body of surrounding ``meta'' interactions, including categories and ranking of content. 
Hence, we also gather metadata about each access by scraping the web content itself, \eg content category and upload date.
In this paper, we look into three aspects of operation. 
First, we inspect the corpus and workload served by the platform (Section~\ref{sec:corpus}). 
Despite its prominence as a video portal, the access logs are dominated by image content, primarily serving thumbnail content.
That said, we find that the majority of bytes served is actually for video content, primarily due to it voluminous nature. 
Video content tends to be relatively short, with subtle variations observed across the categories. 
Popularity across these resources is highly skewed though: the top 10\% of videos contribute 73.7\% of all accesses.
This leads us to explore the specifics of per-session access patterns on the site (Section~\ref{sec:behaviour}). 
We see that, for instance, the majority of sessions limit accesses to one or two categories. 
This leads us to inspect \emph{where} accesses come from. 
The majority of views arrive from  the main video page, but we also observe a number of views from the homepage and search function. 
Finally, we discuss potential implications from our work (Section~\ref{sec:implications}). 
We find that this genre of material is highly cacheable, and briefly test the efficacy of city-wide edge cache deployment. 
We conclude by proposing simple innovations that could streamline delivery (Section~\ref{sec:conclusion}). 
%

\section{Background \& Related Work}

Pornography is amongst the most searched for content on the web~\cite{wondracek10,Ogas11}. 
Although this topic remains a taboo in some research fields, there has been an expanding body of research into the video platforms that drive its delivery.
We have seen recent work inspecting the content corpus of popular adult websites~\cite{tyson2016measurements,tyson2013demystifying} and their workloads~\cite{morichetta2019characterizing,yu2019comparative,ahmed2016internet,zhang2019measurement} as well as various studies that have attempted to estimate the load that they create on the wider Internet. 
For example, \cite{Ogas11} estimates that Porn 2.0 sites such as xHamster and YouPorn can gain up to 16 
million views per month. 
There have also been a number of related studies that have explored the topic of online pornography more generally, \eg privacy~\cite{vallina2019tales}; automated recognition and classification \cite{Mehta97,Hu07}; interest recommendations~\cite{schuhmacher9exploring}; and security issues~\cite{wondracek10}. 
This paper presents one of the first large-scale studies of an online adult multimedia delivery service. 
That said, there are a multitude of studies into more traditional video streaming systems that already provide some insight.
These include catch-up TV \cite{Abrahamsson2012popularity,NRS13www}, user generated content \cite{raman2018facebook,YouTube,zink2009characteristics}, Video-on-Demand \cite{yu2006understanding} and IPTV \cite{cha2008watching,Gao09}.
These insights have been used to drive a range of modeling and systems research activities, \eg building content popularity models \cite{guo2008stretched}, optimized caching techniques \cite{Abrahamsson2012popularity} and improved delivery schemes \cite{apostolopoulos2002video}. 
The paucity of data related to adult video access, however, makes it difficult to appreciate the applicability of these technologies to this particular field. We write this paper to shed insight into the session-level specifics of adult content access patterns.

\section{Methodology and Data}

The work in this paper relies on two key datasets. 
First, we utilize a dataset provided by a CDN, which captures access logs to an anonymous major adult video content provider. 
Second, we compliment this with web metadata surrounding each video in the access logs.

\subsection{CDN Data}

We first describe the basic features of the CDN data, collected in 2019, as well as the necessary post-processing required to extract sessions.

\pb{Data Description.}
We have been given access logs for web resources hosted by a major adult video website. 
The data has been collected from the vantage point of a single US-based data center operated by a major Content Delivery Network (CDN). 
The dataset covers 1 hour, and includes 20.08M access entries.
Each log entry maps to a single resource request, consisting of: 

\begin{itemize}
    \item Timestamp: The time when the item was requested.
    \item Client ID: This is a prefix preserving anonymized IP Address \oliver{(so that we can approximate client location).} 
    \item Resource: The web resource requested.
    \item User Agent: This is the user-agent identifier included within the HTTP request header. This allows us to differentiate mobile from desktop users. 
    \item HTTP Referrer: This is the Referrer from the HTTP request header; it provides the URL that the client was redirected from. 
    \item City ID: Using Maxmind, the anonymized IP addresses are mapped to their geolocation. Only requests for which the coordinates have an estimated accuracy of less than 20KM are tagged. This covers 75.91\% of all requests. Each city is then associated with an anonymized numerical City ID. 
\end{itemize}

\pb{Identifying Sessions.}
For the CDN traces, we take a simple but effective way of mapping the requests to sessions. For each log entry, we generate a session identifier by computing the following hash: \textsf{SHA256(IP Address, Device, Browser)}.
We then group requests into individual sessions using their identifiers. 
Overall, the data covers 62K unique user sessions.
To remove incomplete sessions, we extract all sessions that contain requests within the first or last 5 minutes of the trace, and then filter them from the rest of the logs. 
This removes 15\% of requests. Note that we also performed our analysis on the unfiltered data and only found slight differences.
\subsection{Web Scrape Data}
The CDN logs offer fine-grained insight into individual access patterns, but no metadata related to the content being accessed.
Hence, we scrape metadata from the website front end for each video contained within the CDN dataset. 
The web scrape data, for each video, includes the category, the global view counter, the number of likes/dislikes and any associated hashtags. 
In total we gathered this metadata for 4.9 million videos, covering 91.1\% of all videos in the CDN traces. 

\subsection{Ethical Considerations \& Limitations}

\pb{Limitations}
We emphasize that the duration of our trace data limits our ability to make generalizable statements, particularly pertaining to longitudinal trends.
Critically, this creates a clear bias towards shorter sessions that do not exceed an hour (we filter out longer sessions that do not entirely fall within the measurement period).
Another limitation is that the data only covers a single portal from the vantage point of a single data center. 
Hence, we cannot quantify to what extent this applies to alternative deployments. 
We therefore temper our later analysis with these observations; despite this, we argue that the data offers a powerful first insight into traffic patterns within this domain. 

\pb{Ethical Consideration}
We took a number of steps to address ethical concerns. 
Before receiving the logs, they were first fully anonymized by the CDN, and there was no way to map logs back to specific users.
Hence, all user identifiers were removed prior to access, including cookies and source IP addresses. 
We further anonymized sensitive data (such as content category tags) from the web data, instead generating a set of neutral tags.
Although this restricted ``semantic'' analysis of the content, it reduced exposure to sensitive insight. 
Pre-processing was done by one author, who did not perform the subsequent analysis. 
Furthermore, all data was stored remotely in a secure silo with restricted access to just two authors. We obtained IRB approval.
%

\section{Characterization of Corpus \& Workload}
\label{sec:corpus}

We start by performing a basic characterization of the corpus served, as well as the overall site workloads observed at the CDN.

\begin{figure}
    \centering
    \includegraphics[width=0.45\linewidth]{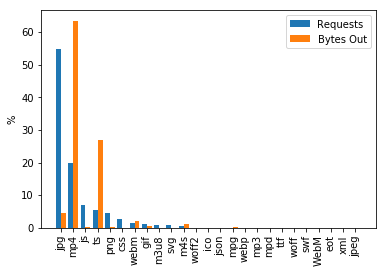}
    \includegraphics[width=0.45\linewidth]{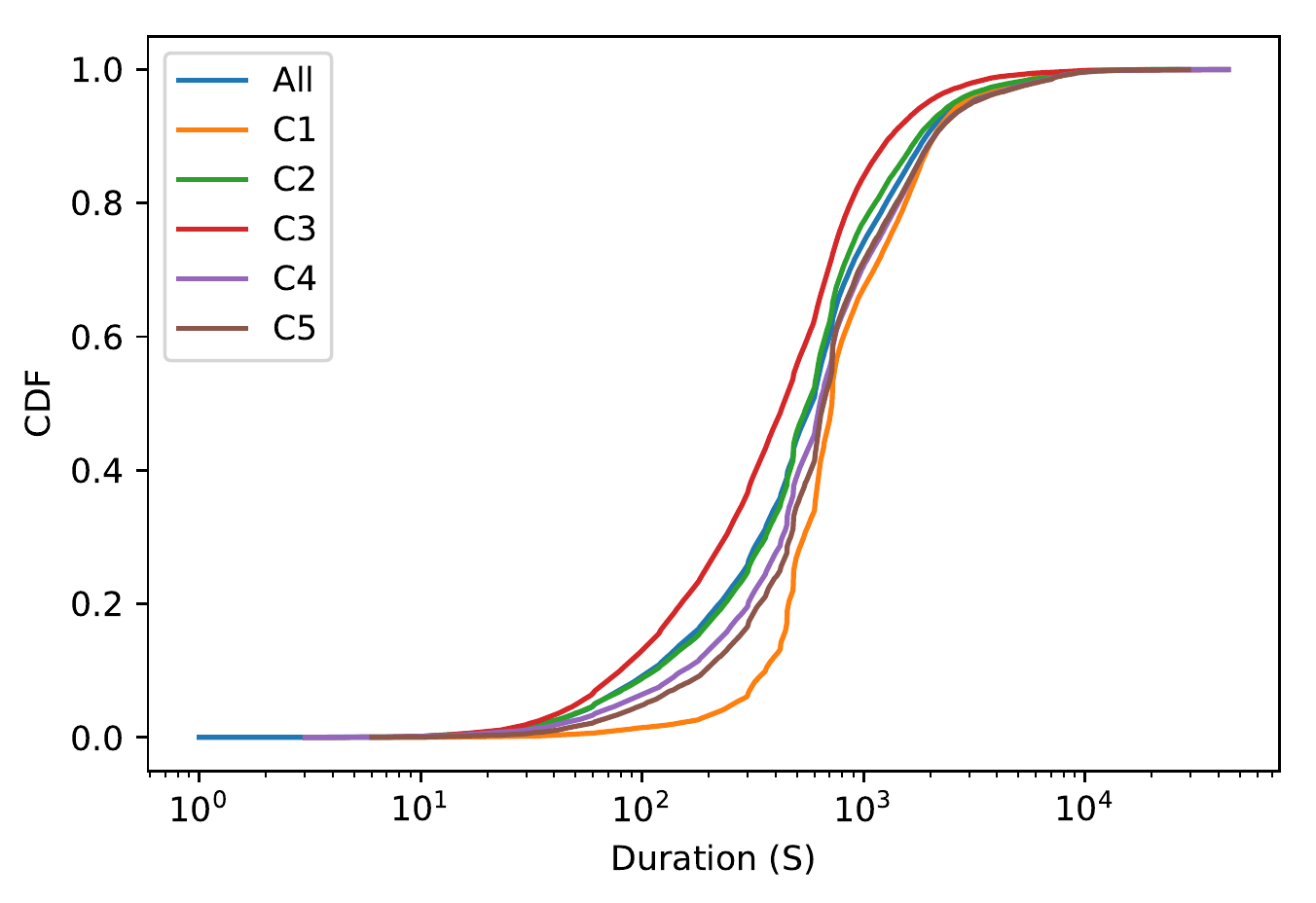}
    \caption{(a) Percentage of requests to various file formats and the percentage of total bytes out; (b) CDF of consumed video duration based on category using all and top-$5$ categories. Note ``All'' refers to all content within any category.}
    \label{fig:file_types}
\end{figure}

\pb{Resource Type.}
Typical sites consist of a wide range of media. 
To inspect this, we first look at the mix of content types encountered within the CDN logs. 
Figure~\ref{fig:file_types}a presents the fraction of requests to each resource type; this shows the distributions for both the number of requests and the number of bytes sent by the servers.
The vast majority of requests are received for image content (mainly jpg), whereas the majority of bytes are attributed to the delivery of video content (mp4). 
In total, 63.4\% of bytes transferred are attributable to video content, yet they constitute only 19.9\% of requests. 
Closer inspection confirms that the dominance of images is driven by thumbnails, which  makes the video portal image-heavy. 
We conjecture that these may be used heavily when browsing, due to the primarily visual nature of the content.

\pb{Video Duration.}
The above suggests that the majority of accesses are actually driven by non-video material. 
Despite this, due to its voluminous nature, the quantity of bytes transferred is dominated by video content. 
Hence, we next inspect the duration of video content available.
We take this from the web scrape information, which includes the video duration. 
Figure~\ref{fig:file_types}b presents the duration of videos on the website, as reported by the video metadata. 
The majority of videos (80\%) fall below 16 minutes, with a mean duration of 920 seconds. 
For completeness, we also plot the duration for videos within the top 5 most popular categories.
Note for ethical reasons, we anonymize all categories.
The shortest videos fall into the C3 category, with a mean of 657 seconds. 
This particular category of material focuses primarily all more homemade content. 
In contrast, the C1 category (which contains more professional material) has a longer mean duration (1086 seconds). 
That said, these categories show strong similarities in their distribution, showing a bias towards shorter content items.
%

\begin{figure}[t]
    \centering
    \includegraphics[width=0.45\linewidth]{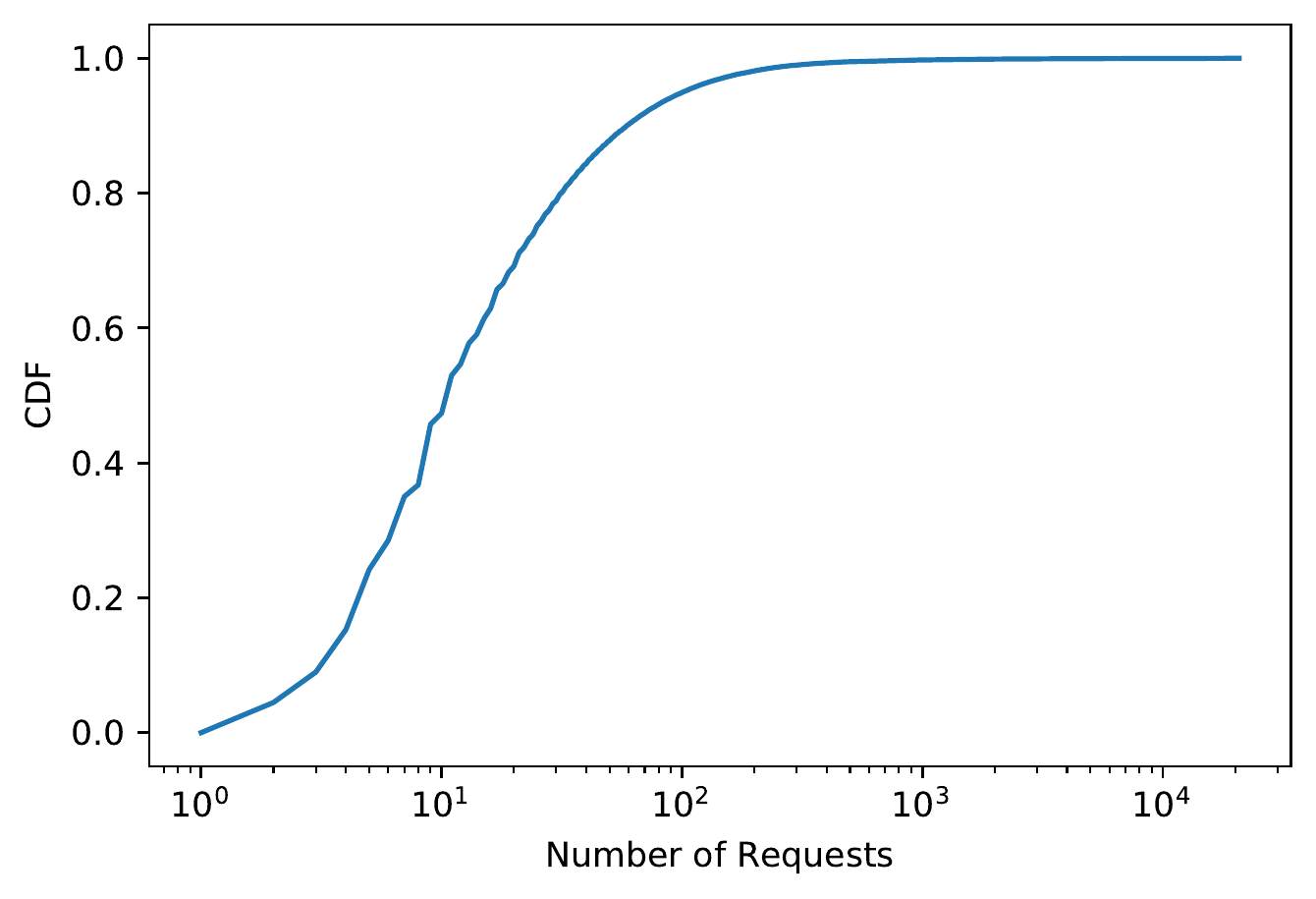}
    \includegraphics[width=0.45\linewidth]{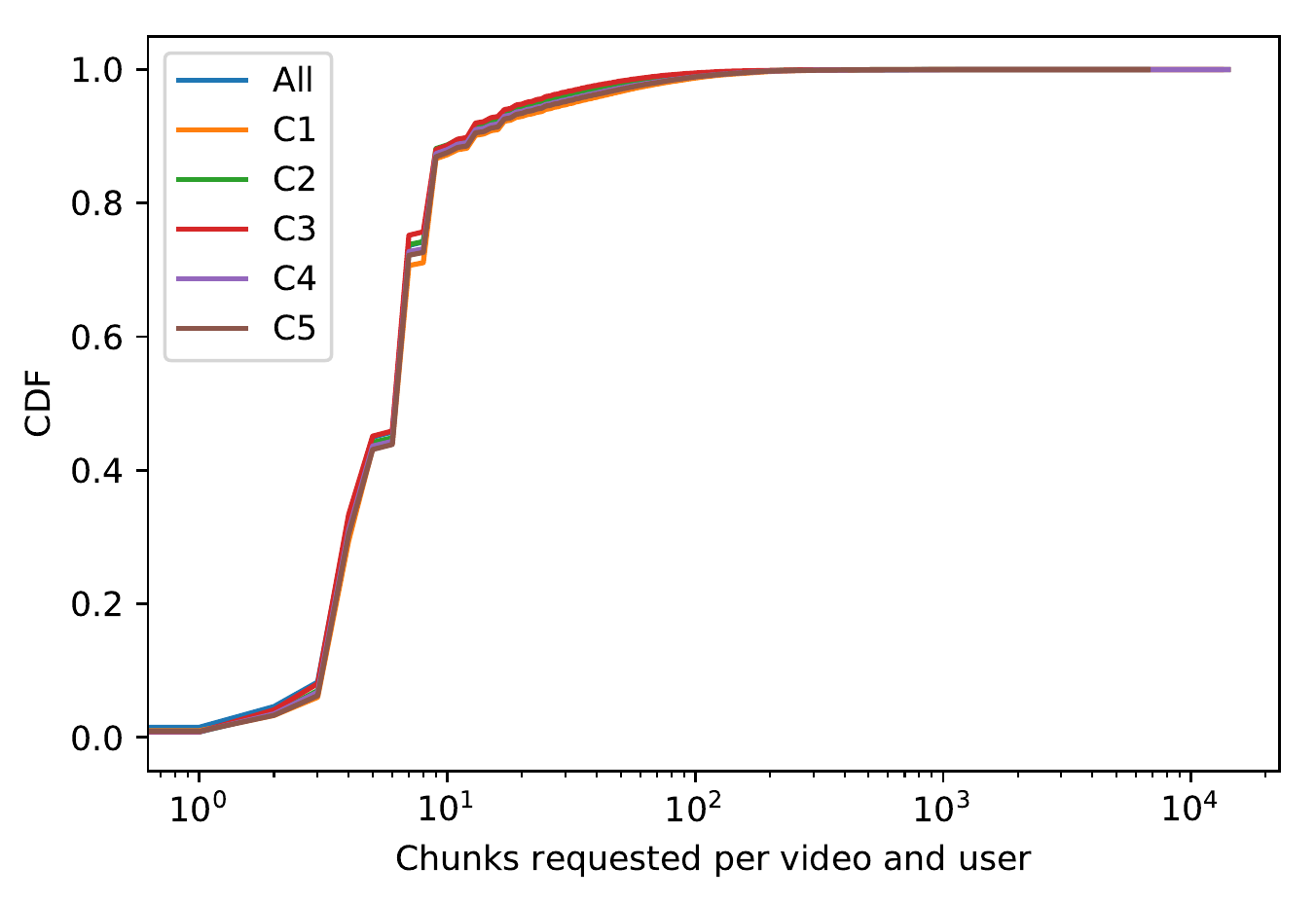}
    \caption{(a) Number of requests per object; 
    (b) Distribution of video chunk per video request} 
    \label{fig:req_per_object}
\end{figure}

\pb{View Counts.}
We next seek to explore the popularity distribution of the resources within our logs.
Figure~\ref{fig:req_per_object}a presents the CDF of the number of requests we observe per-object, taken from the CDN logs.
We observe a clear skew, where the majority of accesses are accumulated by a small number of videos: The top 10\% of videos contribute 73.7\% of all accesses.
This, however, can be deceptive as videos are quite diverse (\eg in terms of duration), and many of the objects downloaded are non-video. 
Hence, Figure~\ref{fig:req_per_object}b complements these results by presenting the CDF of the number of chunks requested per \emph{video}.
Each chunk represents a subset of the overall video content.
This provides insight into the number of sub-requests triggered by each video being consumed: By definition, longer videos will generate more chunk requests. 
Again, we separate chunks into their respective anonymized categories. 
We see that the \emph{vast} majority of video fetches result in under 10 chunks being served.
Initially, one might assume that this is simply because the videos are short. 
However, we find that the low fetch rates are also driven by user \emph{skipping} and \emph{cancellations}, leading to only subsets of a video's chunks be downloaded. We revisit this observation in Section~\ref{sec:behaviour}.

\begin{figure}[t]
    \centering
    \includegraphics[width=0.45\linewidth]{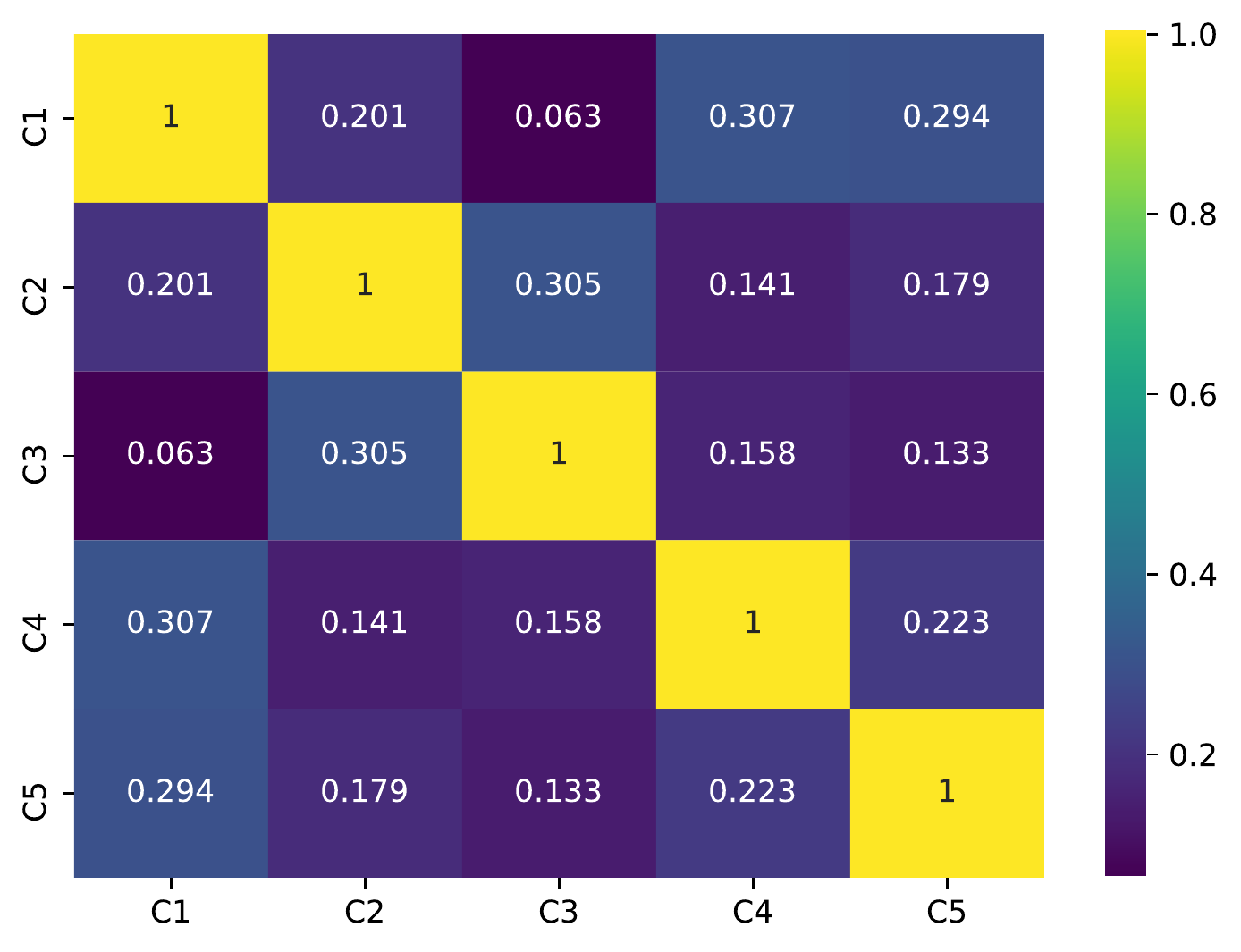}
    \includegraphics[width=0.45\linewidth]{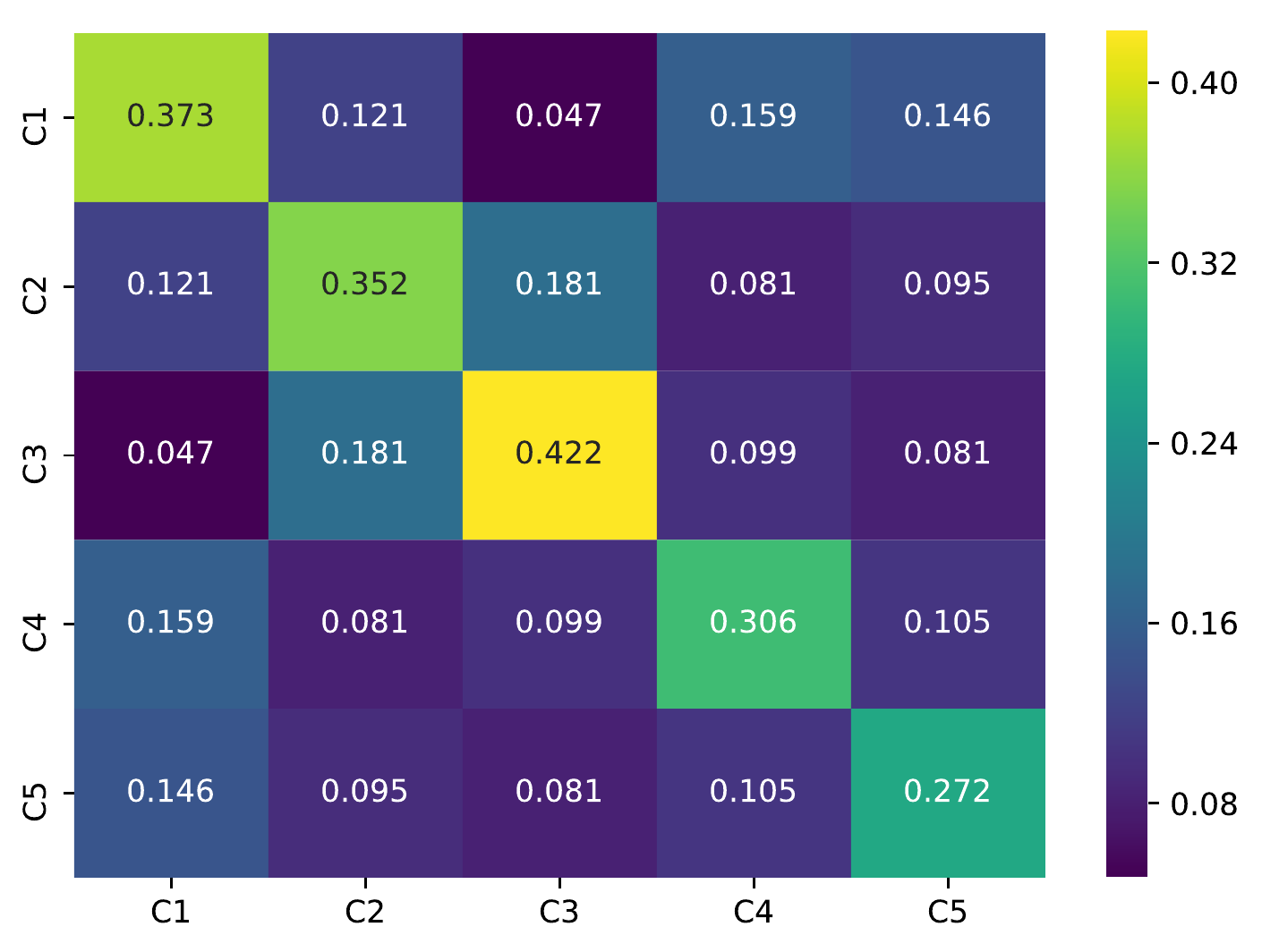}
    \caption{(a) Heatmap showing the fraction of the pair-wise coexistence for the five most popular categories; (b) Heatmap normalised by the total number of videos (across all categories).} 
    \label{fig:heatmap}
\end{figure}

\pb{Category Affinity.}
The above has shown that there are subtle differences between categories of content, \eg in terms of duration. 
A complicating factor is that many videos are tagged by multiple categories. 
On average, each video has 7 category tags. 
Hence, we next briefly test the coexistence between categories to identify commonly paired tags. 
To quantify this, we compute the fraction of the pair-wise coexistence of the top 6 categories and present the results as a heatmap in~\autoref{fig:heatmap}a.
To compute this, we calculate the fraction of videos from each category that \emph{also} are tagged with another category. 
For completeness, ~\autoref{fig:heatmap}b also normalizes the fraction based on the total number of videos.
We confirm that there are varying levels of category co-location. 
In some cases, co-location is quite high, \eg 29.4\% of of videos tagged as C2 are also tagged a C3.
In contrast, other categories are far less co-located, \eg less than 5\% of C1 videos are co-located with C3. 
There are certain intuitive reasons for this, driven by the semantic nature of the categories. We posit that this may offer insight into how things like predictive caching could be introduced to such platforms.

\section{Characterisation of Per-Session Journey}
\label{sec:behaviour}

We have so far revealed a workload dominated by image and video content, as well as patterns which suggest that users rarely consume entire videos. 
Thus, we next proceed to focus on the behavior of individual sessions.

\subsection{Intra-Video Access Journeys.}

We first dive into the \emph{intra}-video access patterns of sessions. 
Our focus is on understanding how users move between chunks within a single video. 

\begin{figure}[t]
    \centering
    \includegraphics[width=0.45\linewidth]{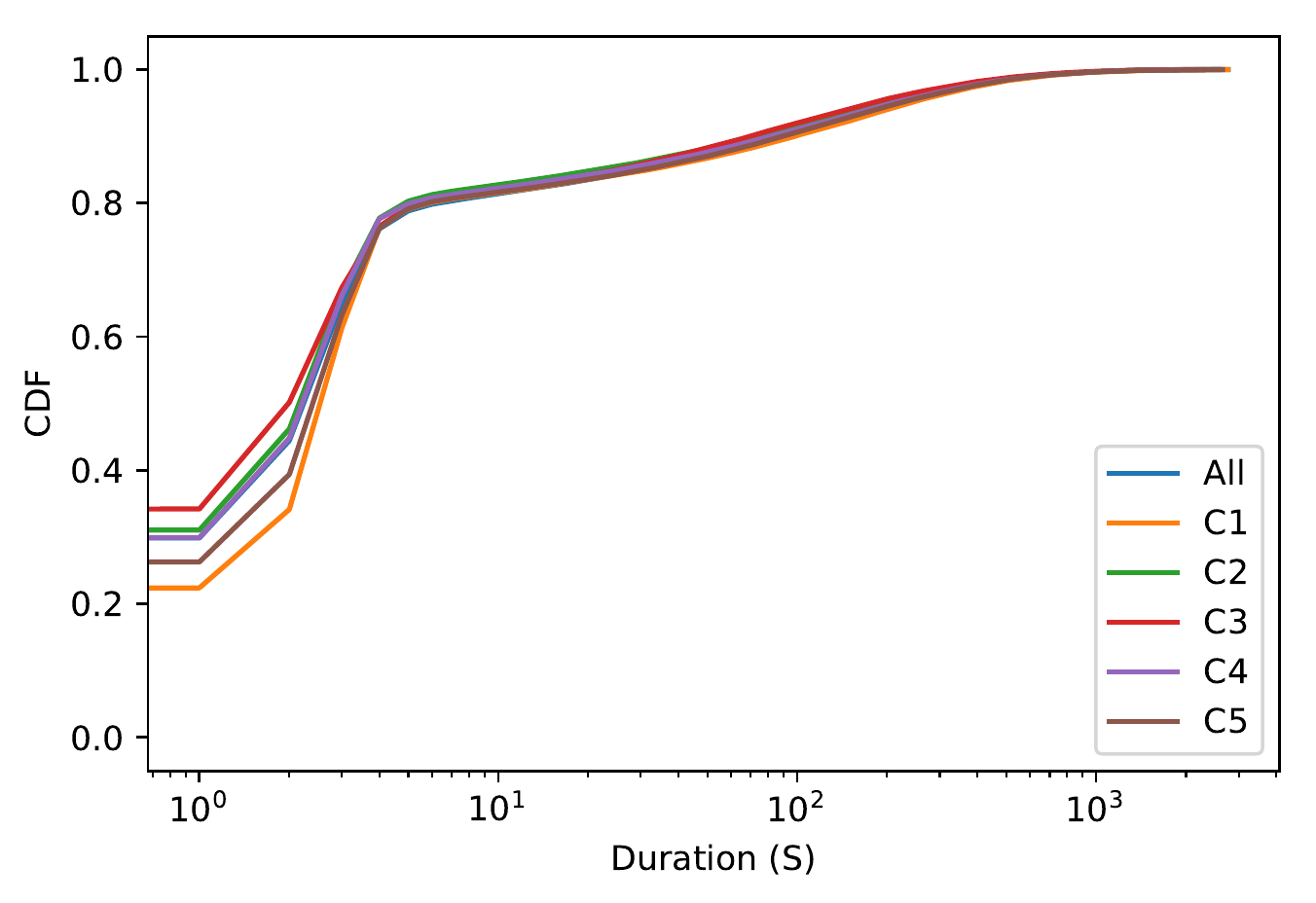}
    \includegraphics[width=0.45\linewidth]{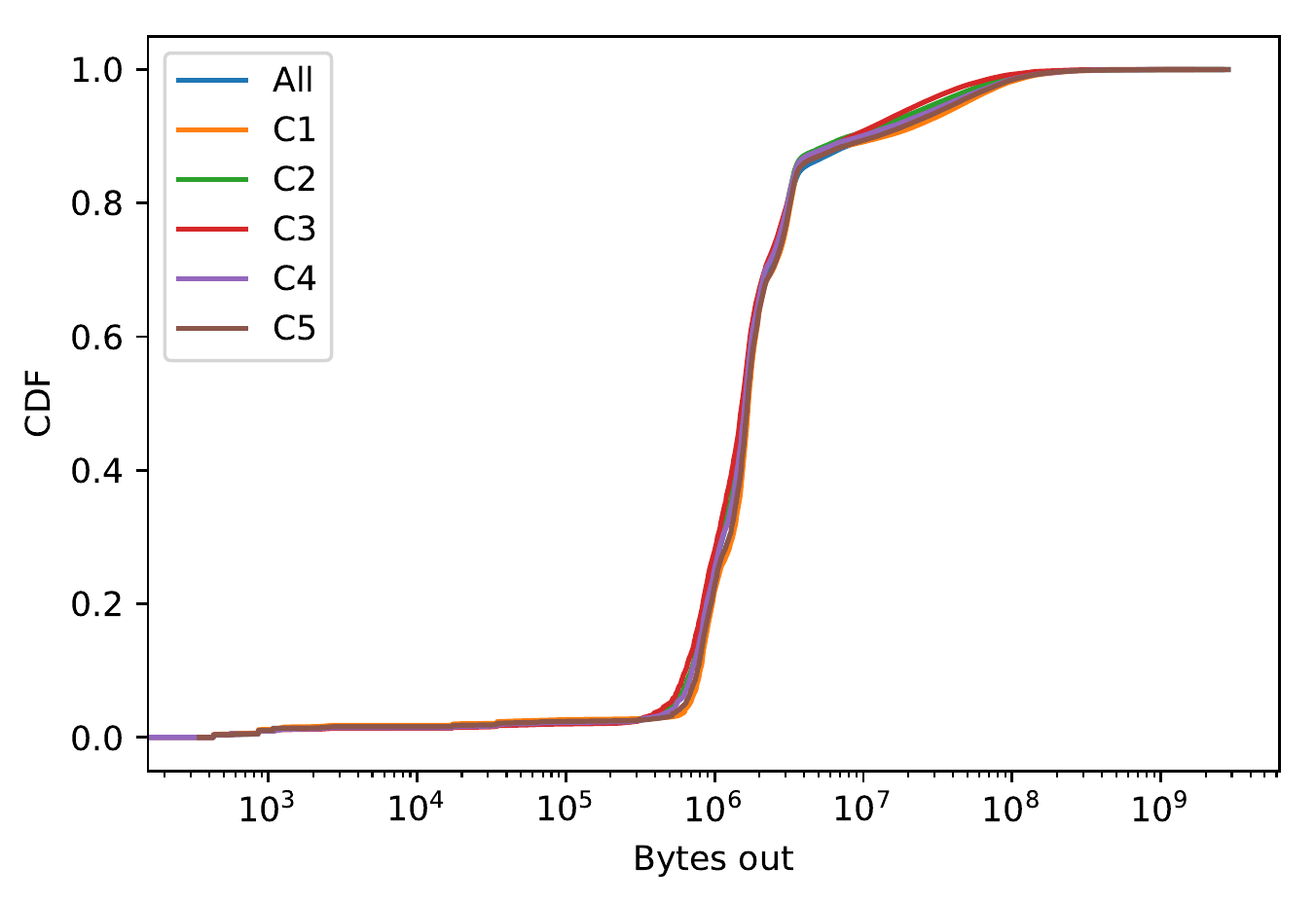}
    \includegraphics[scale=0.45]{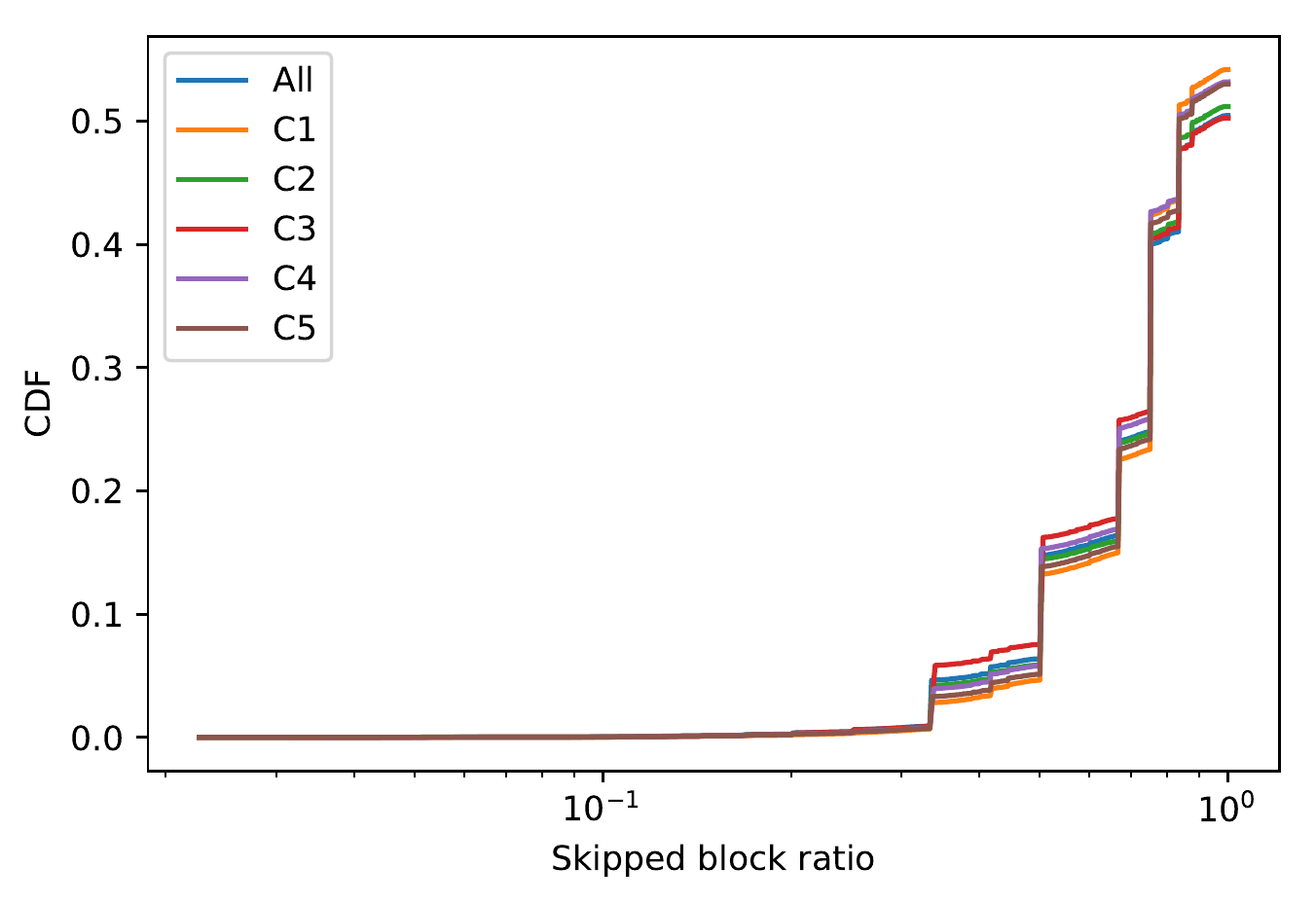}
    \caption{(a) CDF of the approximate consumption for each individual video across sessions for all and top-$5$ categories; (b) CDF of the bytes out per User/Video combination for all and top-$5$ categories (c) Skipped blocks for each category.}
    \label{fig:user_id_grouped_by_timestamp_cats}
\end{figure}

\pb{Access Duration.}
We first explore the duration of time each user session dedicates to an individual video. 
Note that this is different to Figure~\ref{fig:file_types}b, which is based on the video duration, rather than the access duration. 
To compute this, for each video access, we extract the \emph{difference} between the first and last timestamp seen for chunks of the same video.
For instance, if the first chunk of a video were requested at $t^1$, and the final chunk were requested as $t^2$, we estimate the duration as $t^2-t^1$. 
This offers an approximation of access duration, although we highlight that the downloading of a chunk does \emph{not} necessarily mean it is viewed. 
Figure~\ref{fig:user_id_grouped_by_timestamp_cats}a presents the results as a CDF. 
This shows markedly different trends to that of Figure~\ref{fig:file_types}b (which depicts the duration of the content). 
As expected, we find that access durations are far shorter than the underlying content duration that is being consumed. 
There are also subtle differences between the categories; for example, the average access duration for content within the C1 category is 1086 seconds \vs 657 seconds for C3 content. 
Around 80\% of C1 videos are consumed for under 1000 seconds, whereas this is closer to 90\% for C3 videos. 
To complement this, \ref{fig:user_id_grouped_by_timestamp_cats}b presents a CDF of the number of bytes sent per [video, session] pair. 
Each data point represents the number of bytes downloaded for each request (note one session may generate multiple requests, even for the same resource). 
This shows a rather different trends, with the around 90\% of fetches resulting in under $10^7$ bytes being sent. 

Overall, both plots reveal that the majority of videos only have only a subset of their content chunks fetched. 
It is worth noting that, even though videos rarely download all their chunks, we do find that requests for individual chunks are usually completed. 82\% of individual chunk requests involve downloading in excess of 90\% of bytes, whilst only 4\% download under 10\% of bytes.

\pb{Cancellations and Skip Rates.}
The fact that many videos are not downloaded in their entirety is driven by a combination of two factors: \one~viewers canceling video streams; and \two~viewers skipping across video streams.

To get an idea of how many videos are watched sequentially, and then canceled towards the end, we compute the fraction of streams that request the first 90\% of chunks, but lack the last 10\%. We find that under 1\% experience this, suggesting that early cancellations and skips are most prevalent. 
Figure~\ref{fig:user_id_grouped_by_timestamp_cats}c presents the skip rate of blocks. 
A skipped block is counted when the byte range is not directly adjacent to the previous block high range. 
For example, a contiguous block is: ``100-200'' and ``201-300'', whereas a skipped block is ``100-200'' and then ``501-600''.
We observe that some videos have extremely high skip ratios (\ie above 0.8). 
This confirms that viewers skip extensively within videos, and rarely download all chunks contiguously. 
This has a dramatic impact on our earlier results. 
To quantify this, we subset all videos to leave those containing at least one skip (and remove any anomalous blocks as mentioned previously).
This leaves all videos served that have \textit{at least one skip} --- this covers a 75.4\% of the total requests, confirming that the majority of videos do include skips. 
This is likely to differ from long-play Video-on-Demand platforms (\eg Netflix) where users more likely view streams contiguously.
%

\subsection{Inter-Video Access Journeys}

The next aspect we inspect is how sessions move \emph{between} videos. 

\begin{figure}[t]
    \centering
    \includegraphics[width=0.45\linewidth]{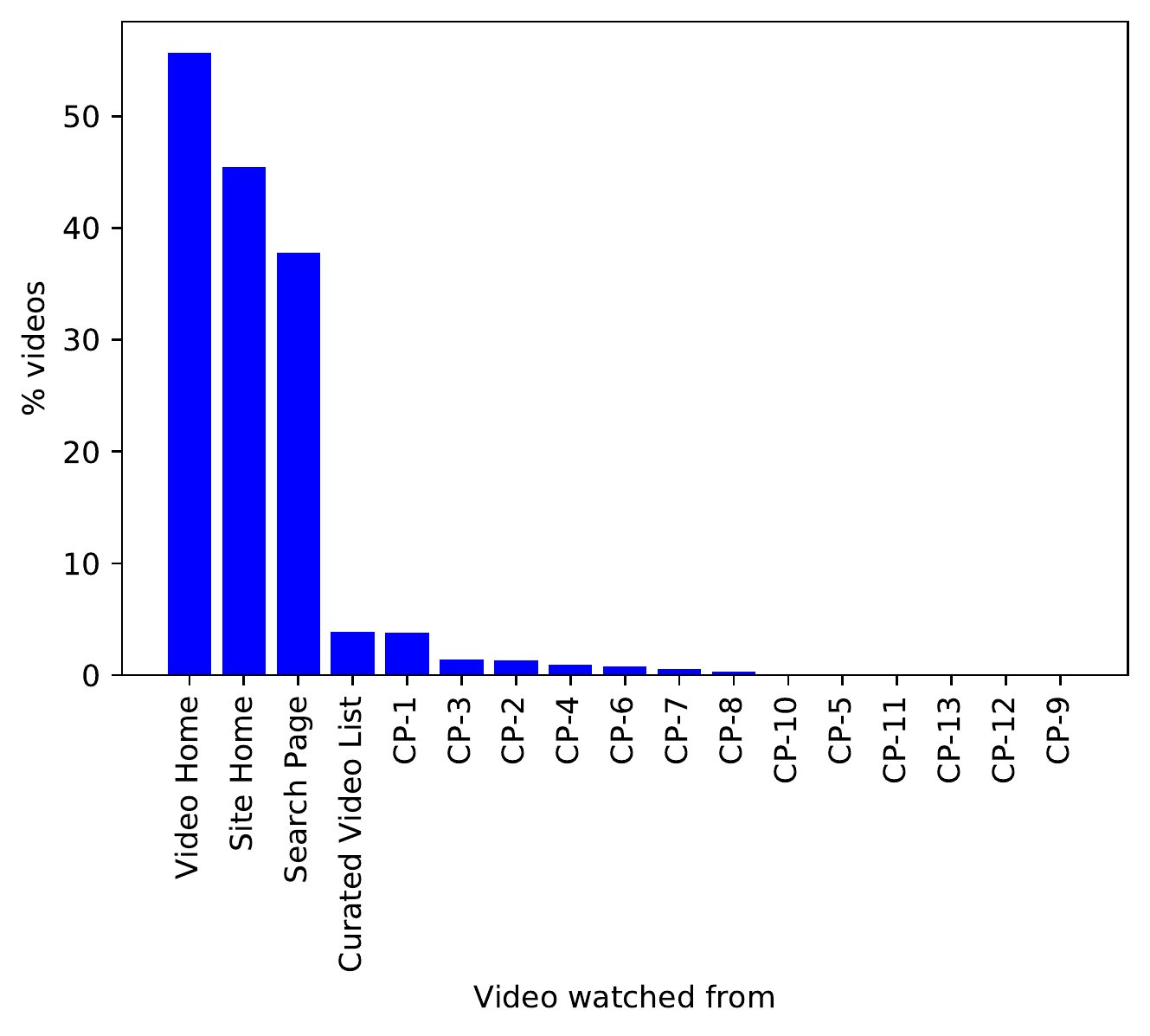}
     \includegraphics[width=0.45\linewidth]{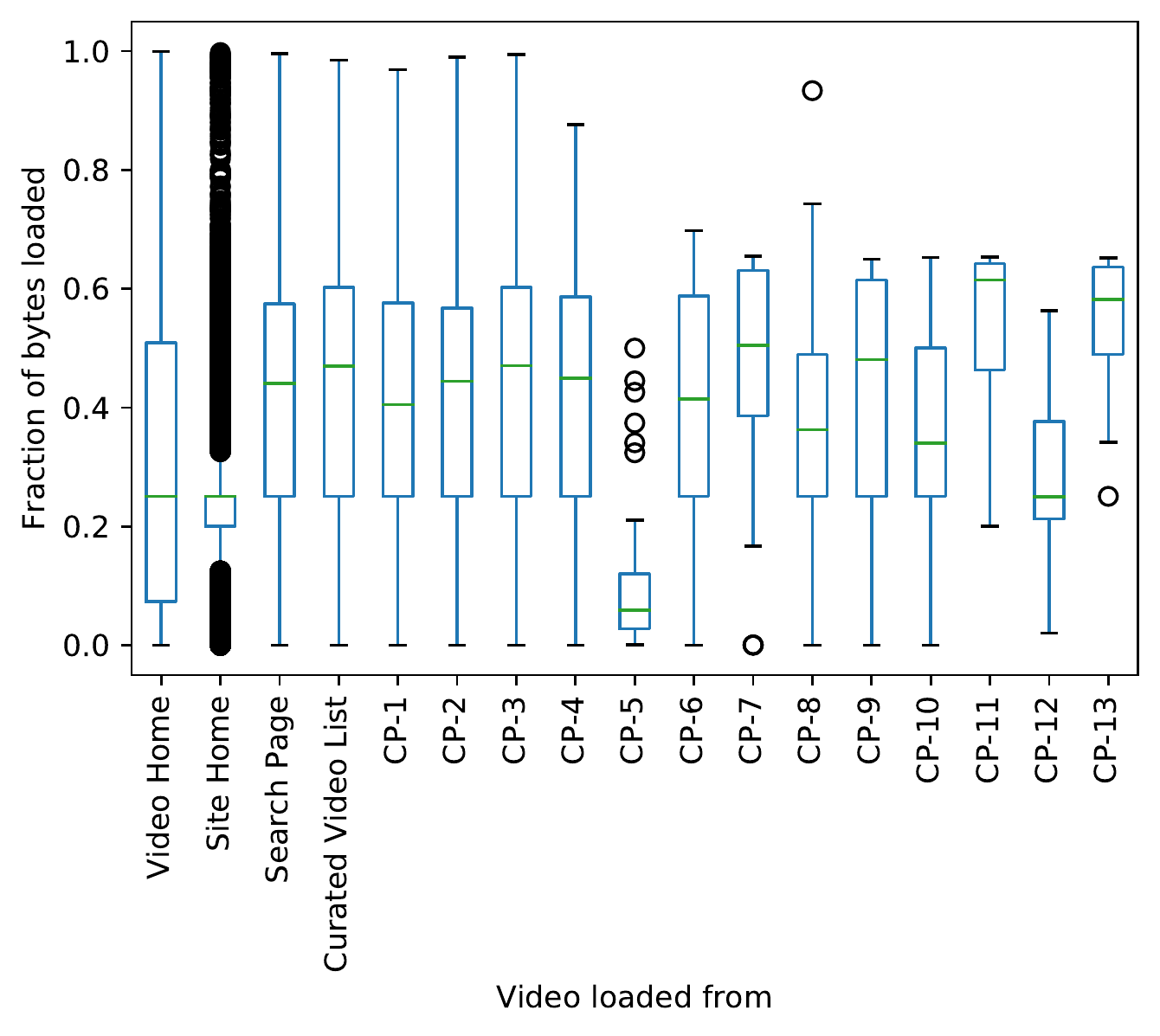}
    \caption{(a) Where the videos are watched the most: 95.67\% of videos are watched from either the main page of the video, the homepage of the site and the search page (b) Where the videos are loaded the most: Y-axis gives the ratio of bytes out and total file size (somewhat indicating what proportion of video has been watched) across users from various pages }
    \label{fig:video_refs}
\end{figure}

\pb{Video Load Points.}
We first inspect which pages tend to drive the majority of video views. 
We conjecture that different viewers might have different patterns in this regard. 
To extract this information, we identify the HTTP Referrer in each request; from this, we take the previous page the resources was loaded from. 
We then map this to the page and type of object that has triggered the resource request.
Figure~\ref{fig:video_refs}a presents the overall distribution of videos watched from a page that users are visiting within the portal. 
Note that we anonymize category pages again.
The majority of resources are watched from the Video Homepage (each video has its own page). 
This captures over 55\% of unique videos accumulating 65.5\% of bytes delivered on the site.
That said, we also observe a notable quantity of material embedded within the Site Homepage and from the Search Page. 
For instance, around 45.5\% of video visits come from Site Homepage.
Interestingly 37\% of the videos are referred from the Search Page but amassing just 5\% of the traffic. 
The remaining referrals are from various sub-pages within the site, most notably several popular category pages.
Looking at this distribution in isolation, however, is insufficient to gain vantage into a sessions journey. 
This is because, as previously observed, videos are not always viewed in their entirety. 
To explore this further, Figure~\ref{fig:video_refs}b presents the fraction of bytes loaded across the various referrers previously discussed.
For clarity, we list only the top pages observed. 
The median is relatively stable across most pages, however, there are key differences. 
For example, 45.69\% of views from the homepage of the site result in under 25\% of video bytes actually being loaded. 
This might indicate that content accessed from the front page is rarely done with great thought. 
Rather, users might informally click on videos on the chance that they might be of interest. 
Similarly, just 5\% of video bytes are consumed when redirected from the search page, suggesting that users may load a large number of videos in the hope of finding a specific one (before canceling). We will seek to verify these conjectures in our future work. 

\begin{figure}[!ht]
    \centering
    \includegraphics[width=0.85\linewidth]{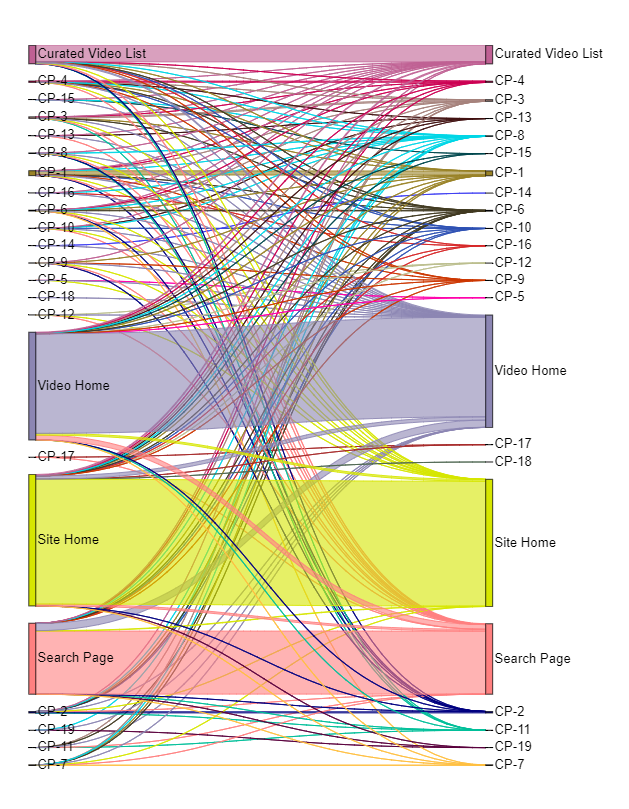}
    \caption{Sankey diagram presenting the fraction of page transitions from locations (left) to destinations (right). This is computed by computing the time ordered list of resources and checking the previous resource request to determine the step-by-step journey.}
    \label{fig:sankey}
\end{figure}

\pb{Inter Video Navigation.}
Whereas the above inspects the number of videos loaded from a given page, it is also interesting to explore the transition of views between videos.
To compute this, we sort each user session into its temporal order of access. This only covers video accesses. We then compute transition points as the move between one resource request to the next. 
Figure~\ref{fig:sankey} presents a Sankey diagram to reveal the transition of accesses between videos. 
We find that the majority of sessions move between resources on the same page type. 
For example, 92.6\% of accesses to the homepage of a video are followed by another video access from the homepage. 
This observation generalizes across most pages. 
For the top 5 accessed pages, we find at least 88.83\% of videos are accessed from the same source as the previous video.
We conjecture that this may be a powerful observation for performing predictive pre-fetching of content.

\section{Discussion \& Implications}
\label{sec:implications}

There are a number of implications of our work. 
Here we briefly focus on potential work in relation to optimizing CDN delivery. 

\pb{Geo-Aware Caching.}
CDNs are primarily interested in improving their quality of provision, as well as overheads. 
This is typically measured via metrics such as cache hit rate \vs deployment costs. 
Our results confirm that, even though images constitute the bulk of requests, the majority of bytes delivered are video content. 
Furthermore, due to the presence of highly popular objects, we posit that there may be potential for edge caching of content.
Although CDNs already deploy cache servers around the world, we next test the possibility of deploying a larger number of \emph{geo-aware} caches.
As we do not have topological information about clients, we cluster users into a cache domain based on their city tags derived from Maxmind.
Note that this creates a wider dispersal of cache servers compared to most CDNs~\cite{ager2011web}.
We then sub-divide users into their cities, and filter any cities that have 10 or fewer sessions, leaving 385 cities. 
For simplicity, we assign all users in each city into a single caching domain, assuming that each region has its own dedicated geo-cache.

\begin{figure}[t]
    \centering
    \includegraphics[width=0.45\linewidth]{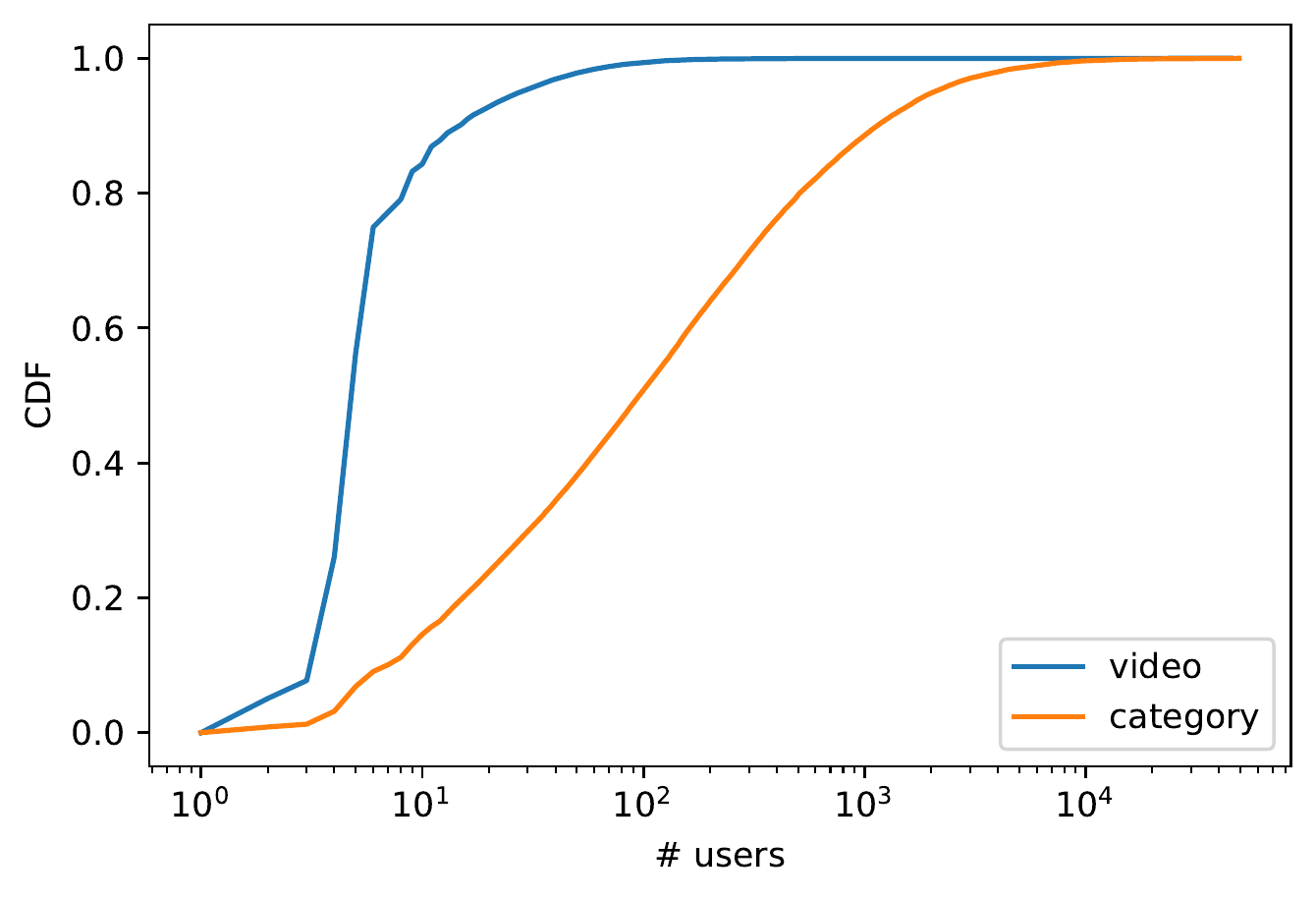}
    \includegraphics[width=0.5\linewidth]{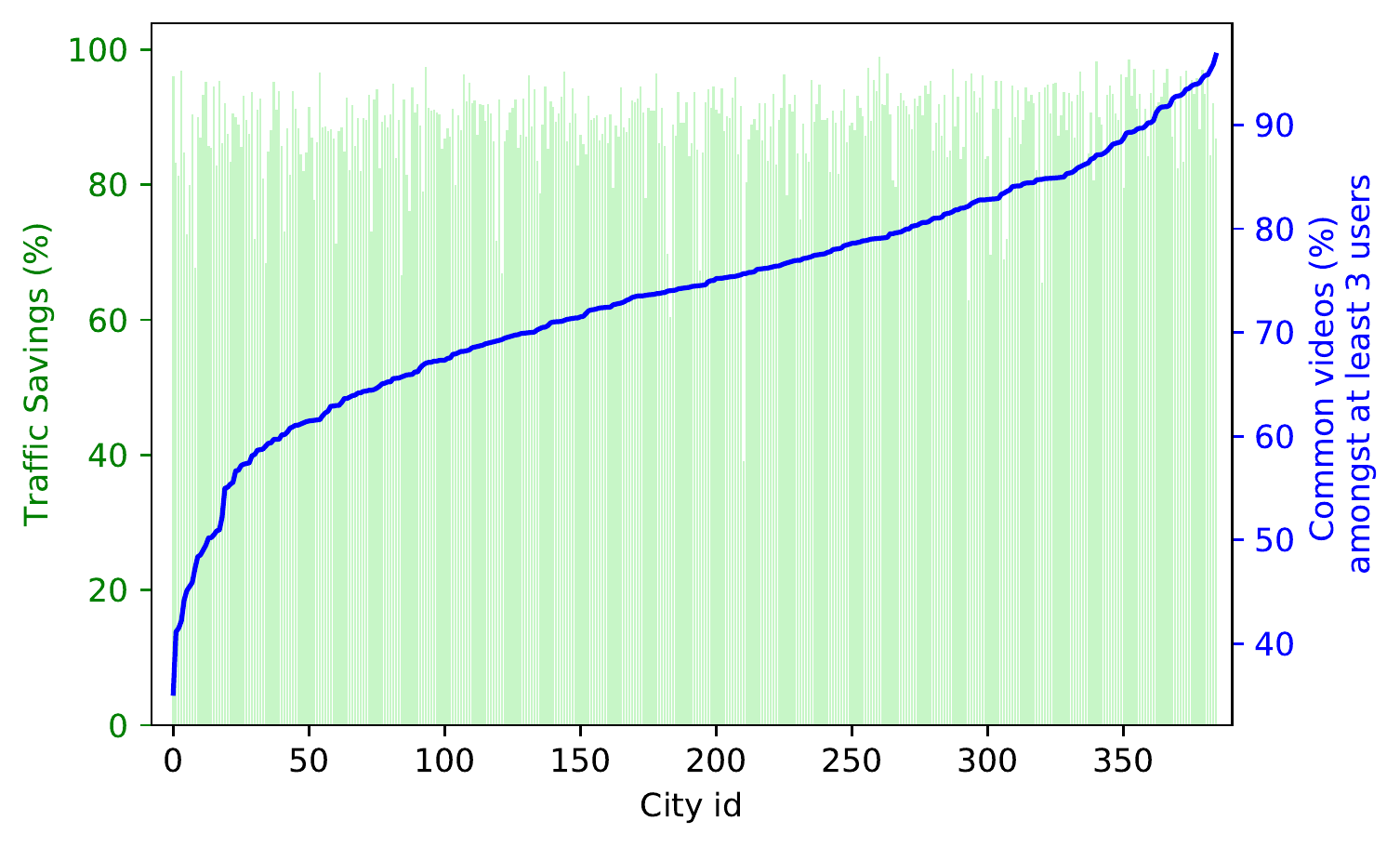}
    \caption{(a) CDF of number of users who have watched the same video in their city (blue) or a video from the same category in their city (orange); (b) Percentage of traffic saved at back-haul by implementing city-wide cache (Y-1) and the percentage of users who would have benefit by the scheme (Y-2).} 
    \label{fig:geo_similarity}
\end{figure}

We first compute how many sessions in each city consume the same video. 
Figure~\ref{fig:geo_similarity}(a) presents the results on a per-video basis. 
Unsurprisingly, we find that accessing the same video from a city is commonplace. 
In the most extreme case, one video is accessed by 98.9\% of all sessions within a particular city. 
This leads us to hypothesize that such properties could be exploited for caching. 
Hence, Figure~\ref{fig:geo_similarity}(b) shows the percentage of traffic that could be saved (Y-1 axis) \emph{if} a city-wide cache were to be deployed. 
Note, for simplicity, we assume the cache covers all users in the city and has unlimited storage for the one hour period of the dataset. 
For these high population locations, savings exceeding 90\% are feasible. 
The Y-2 axis also presents the percentage of videos that have at least 3 user sessions within a city accessing them (\ie thereby resulting in a saving). 
We see that these are extremely high, with nearly all cities exceeding 50\%. 

\pb{Predictive Loading.}
The above confirms that caching is an effective tool in this domain. 
We also posit that a number of more innovative approaches could be taken for streamlining delivery. 
For instance, \emph{predicting} popular chunks in the video and subsequently pushing them could improve Quality of Experience by reducing human-perceived delays. 
This would be particularly useful, as often videos are not viewed contiguously, making current buffering strategies ineffective. 
Predicting the next skip could therefore avoid wasted buffering. 
Furthermore, the heavy load created by thumbnails, suggest they could perhaps be pre-loaded in bulk for certain videos
We have also confirmed that sessions have clear behavioral traits when moving \emph{between} video pages. 
Again, we conjecture that these patterns could be predicted and exploited. 
For instance, the top video within a recommendation pane could be pre-loaded, much as we see done with Accelerated Mobile Pages~\cite{phokeer2019potential}. 
In fact, due to the propensity for viewers to select such content, it might even be possible to dynamically select which videos to recommend based on what content is live in the most nearby cache. 
We posit that this may be able to satisfy user demand, whilst also reducing network costs for the CDN. 
%

\section{Conclusion}
\label{sec:conclusion}
This paper has explored the characteristics of a large adult video portal, with a focus on understanding in-session journeys.
We first inspected the corpus and workload served by our vantage point. 
We found that, contrary to expectation, the bulk of objects served are actually image content, although video \emph{does} make up the bulk of bytes delivered. 
In terms of videos, the majority of requests were for a small subset of the content, and we confirmed past observations related to the skewed distribution of adult content. 
This led us to focus on session-level behaviors, where we revealed distinct access patterns and briefly evaluated the potential of caching and pre-fetching to optimize delivery.
The work constitutes just the first step in our research agenda. 
We have so far studied the journey patterns within sessions, however, we wish to better understand \emph{why} these patterns emerge. 
This generalizes beyond adult video to any type of website. 
Thus, we wish to do further comparative research with other portals.
With these patterns, we also wish to develop optimized delivery systems that can learn behavior sufficiently well to predict and pre-load content per-user. 
Finally, we are keen to deep dive into the innovations discussed, and perform further experiments to understand how they can streamline delivery. 
%

\section*{Acknowledgments}
\label{sec:acknowledgments}

%
This work was supported by EPSRC grants EP/N510129/1 and EP/P025374/1. 
We would also like to thank the reviewers and our shepherd Oliver Hohlfeld.

\bibliographystyle{splncs04} 
\bibliography{refs} 

\begin{thebibliography}{10}
\providecommand{\url}[1]{\texttt{#1}}
\providecommand{\urlprefix}{URL }
\providecommand{\doi}[1]{https://doi.org/#1}

\bibitem{Abrahamsson2012popularity}
Abrahamsson, H., Nordmark, M.: Program popularity and viewer behaviour in a
  large tv-on-demand system. In: Proc. IMC (2012)

\bibitem{ager2011web}
Ager, B., M{\"u}hlbauer, W., Smaragdakis, G., Uhlig, S.: Web content
  cartography. In: Proceedings of the 2011 ACM SIGCOMM conference on Internet
  measurement conference. pp. 585--600 (2011)

\bibitem{ahmed2016internet}
Ahmed, F., Shafiq, M.Z., Liu, A.X.: The internet is for porn: Measurement and
  analysis of online adult traffic. In: 2016 IEEE 36th International Conference
  on Distributed Computing Systems (ICDCS). pp. 88--97. IEEE (2016)

\bibitem{apostolopoulos2002video}
Apostolopoulos, J.G., Tan, W.t., Wee, S.J.: Video streaming: Concepts,
  algorithms, and systems. HP Laboratories, report HPL-2002-260  (2002)

\bibitem{YouTube}
Cha, M., Kwak, H., Rodriguez, P., Ahn, Y.Y., Moon, S.: {Analyzing the Video
  Popularity Characteristics of Large-scale User Generated Content Systems}.
  IEEE/ACM Trans.\ Netw.  \textbf{17}(5),  1357--1370 (2009)

\bibitem{cha2008watching}
Cha, M., Rodriguez, P., Crowcroft, J., Moon, S., Amatriain, X.: Watching
  television over an ip network. In: Proc. IMC. pp. 71--84. ACM (2008)

\bibitem{Gao09}
Gao, P., Liu, T., Chen, Y., Wu, X., Elkhatib, Y., Edwards, C.: The measurement
  and modeling of a p2p streaming video service. Networks for Grid Applications
   \textbf{2},  24--34 (2009)

\bibitem{guo2008stretched}
Guo, L., Tan, E., Chen, S., Xiao, Z., Zhang, X.: The stretched exponential
  distribution of internet media access patterns. In: Proc. PODC. pp. 283--294.
  ACM (2008)

\bibitem{Hu07}
Hu, W., Wu, O., Chen, Z., Fu, Z., Maybank, S.: Recognition of pornographic web
  pages by classifying texts and images. IEEE Trans.\ Pattern Analysis and
  Machine Intelligence  \textbf{29}(6) (2007)

\bibitem{arbor}
Labovitz, C., Lekel-Johnson, S., McPherson, D., Oberheide, J., Jahanian, F.:
  {Internet Inter-Domain Traffic}. In: Proc. SIGCOMM (2010)

\bibitem{Mehta97}
Mehta, M.D., Plaza, D.: Content analysis of pornographic images available on
  the internet. The Information Society  \textbf{13}(2),  153--161 (1997)

\bibitem{morichetta2019characterizing}
Morichetta, A., Trevisan, M., Vassio, L.: Characterizing web pornography
  consumption from passive measurements. In: International Conference on
  Passive and Active Network Measurement (2019)

\bibitem{NRS13www}
Nencioni, G., Sastry, N., Chandaria, J., Crowcroft, J.: Understanding and
  decreasing the network footprint of over-the-top on-demand delivery of tv
  content. In: Proc. World Wide Web Conference (May 2013)

\bibitem{Ogas11}
Ogas, O., Gaddam, S.: A billion wicked thoughts: what the world's largest
  experiment reveals about human desire. Dutton (2011)

\bibitem{phokeer2019potential}
Phokeer, A., Chavula, J., et~al: {On the potential of Google AMP to promote
  local content in developing regions}. In: 2019 11th International Conference
  on Communication Systems \& Networks (COMSNETS) (2019)

\bibitem{raman2018facebook}
Raman, A., Tyson, G., Sastry, N.: {Facebook (A) Live? Are Live Social
  Broadcasts Really Broad casts?} In: Proceedings of the 2018 world wide web
  conference. pp. 1491--1500 (2018)

\bibitem{schuhmacher9exploring}
Schuhmacher, M., Zirn, C., V{\"o}lker, J.: Exploring youporn categories, tags,
  and nicknames for pleasant recommendations. In: Proc. Workshop on Search and
  Exploration of X-Rated Information (SEXI 2013). pp. 27--28 (February 2013)

\bibitem{tyson2013demystifying}
Tyson, G., Elkhatib, Y., Sastry, N., Uhlig, S.: Demystifying porn 2.0: A look
  into a major adult video streaming website. In: Proceedings of the 2013
  conference on Internet measurement conference. pp. 417--426. ACM (2013)

\bibitem{tyson2015people}
Tyson, G., Elkhatib, Y., Sastry, N., Uhlig, S.: Are people really social in
  porn 2.0? In: Ninth International AAAI Conference on Web and Social Media
  (2015)

\bibitem{tyson2016measurements}
Tyson, G., Elkhatib, Y., Sastry, N., Uhlig, S.: Measurements and analysis of a
  major adult video portal. ACM Transactions on Multimedia Computing,
  Communications, and Applications (TOMM)  \textbf{12}(2), ~35 (2016)

\bibitem{vallina2019tales}
Vallina, P., Feal, {\'A}., Gamba, J., Vallina-Rodriguez, N., Anta, A.F.: Tales
  from the porn: A comprehensive privacy analysis of the web porn ecosystem.
  In: Proceedings of the Internet Measurement Conference. pp. 245--258 (2019)

\bibitem{wondracek10}
Wondracek, G., Holz, T., Platzer, C., Kirda, E., Kruegel, C.: Is the internet
  for porn? an insight into the online adult industry. In: Proc. Workshop on
  Economics of Information Security (2010)

\bibitem{yu2006understanding}
Yu, H., Zheng, D., Zhao, B.Y., Zheng, W.: Understanding user behavior in
  large-scale video-on-demand systems. In: ACM SIGOPS Operating Systems Review.
  vol.~40, pp. 333--344. ACM (2006)

\bibitem{yu2019comparative}
Yu, R., Christophersen, C., Song, Y.D., Mahanti, A.: Comparative analysis of
  adult video streaming services: Characteristics and workload. In: 2019
  Network Traffic Measurement and Analysis Conference (TMA). pp. 49--56. IEEE
  (2019)

\bibitem{zhang2019measurement}
Zhang, S., Zhang, H., Yang, J., Song, G., Wu, J.: {Measurement and Analysis of
  Adult Websites in IPv6 Networks}. In: 2019 20th Asia-Pacific Network
  Operations and Management Symposium (APNOMS). pp.~1--6. IEEE (2019)

\bibitem{zink2009characteristics}
Zink, M., Suh, K., Gu, Y., Kurose, J.: Characteristics of youtube network
  traffic at a campus network--measurements, models, and implications. Computer
  Networks  \textbf{53}(4),  501--514 (2009)

\end{thebibliography}

\end{document}